# The upper critical field of filamentary Nb$_3$Sn conductors

A. Godeke

*Applied Superconductivity Center, University of Wisconsin, 1500 Eng. Drive, Madison, WI 53706, USA
and Low Temperature Division, Faculty of Science and Technology, University of Twente, P.O. Box 217, 7500AE Enschede, The Netherlands*

M.C. Jewell, C.M. Fischer, A.A. Squitieri, P.J. Lee and D.C. Larbalestier

*Applied Superconductivity Center, University of Wisconsin, 1500 Eng. Drive, Madison, WI 53706, USA*

We have examined the upper critical field of a large and representative set of present multi-filamentary Nb$_3$Sn wires and one bulk sample over a temperature range from 1.4 K up to the zero field critical temperature. Since all present wires use a solid-state diffusion reaction to form the A15 layers, inhomogeneities with respect to Sn content are inevitable, in contrast to some previously studied homogeneous samples. Our study emphasizes the effects that these inevitable inhomogeneities have on the field-temperature phase boundary. The property inhomogeneities are extracted from field-dependent resistive transitions which we find broaden with increasing inhomogeneity. The upper 90-99 % of the transitions clearly separates alloyed and binary wires but a pure, Cu-free binary bulk sample also exhibits a zero temperature critical field that is comparable to the ternary wires. The highest $\mu_0 H_{c2}$ detected in the ternary wires are remarkably constant: The highest zero temperature upper critical fields and zero field critical temperatures fall within $29.5 \pm 0.3$ T and $17.8 \pm 0.3$ K respectively, independent of the wire layout. The complete field-temperature phase boundary can be described very well with the relatively simple Maki-DeGennes model using a two parameter fit, independent of composition, strain state, sample layout or applied critical state criterion.

## I. INTRODUCTION

The upper critical field ($\mu_0 H_{c2}$) of Nb$_3$Sn is strongly dependent on A15 composition[1]. Although data covering the complete field-temperature (*H-T*) phase boundary are available for one set of pure Nb-Sn thin films[1,2] and a few single- and poly-crystals[3,4], data on practical, inhomogeneous conductors are sparse. Wires may differ substantially in *H-T* behavior compared to homogeneous films and crystals due to inherent inhomogeneities, arising from their manufacture. They exhibit Sn gradients resulting from the solid state diffusion process used for the formation of the A15 layers, and strain gradients caused by lattice imperfections and thermal contraction differences between the composite materials. The Sn gradients originate from the broad composition range of the A15 phase (~18 to ~25 at. % Sn[5]). The existence of these inhomogeneities as well as their potential influence on the superconducting properties have been recognized for a long time[6-13], as is the presence of a scaling field for the critical current which lies substantially below $\mu_0 H_{c2}$. This scaling field is often referred to as $B_{c2}^*$ but in fact it is an irreversibility field ($\mu_0 H^*$) defined by a pinning force or a Kramer function extrapolation[14] ($\mu_0 H_K$). However, to our knowledge no systematic analysis over the complete *H-T* range of the influence of inhomogeneities on $\mu_0 H_{c2}(T)$ and $\mu_0 H^*(T)$ is available in the literature. Moreover, the critical parameters are often defined at an arbitrary criterion of e.g. 50 % or 90 % of the superconducting to normal transition and the behavior at practically important lower temperatures and higher fields is mostly extrapolated from the slope at $T_c(0)$. In our analysis we concentrate on the entire transition widths in an attempt to extract the range of property distributions that are present in wires and measure these over the full relevant *H-T* range.

For wires, $\mu_0 H_{c2}(T)$ data are available in limited fields[15-20] while a few publications report single high field points at liquid helium temperature[17,18,21-23]. Extrapolation of data measured up to 22 T indicates a zero temperature upper critical field ($\mu_0 H_{c2}(0)$) in thin films ranging from 26 to 29 T, depending on resistivity[1,2]. Single- and poly-crystal data, measured over the complete field range, indicate either $\mu_0 H_{c2}(0) \approx 24.5$ T for the tetragonal phase or 29 T for the cubic phase[3,4]. The limited amount of wire $\mu_0 H_{c2}(0)$ data, mostly extrapolated from 4.2 K, ranges from about 20 T up to >30 T[17,18,21-23]. This large spread causes confusion as to what value is to be regarded as realistic in multi-filamentary conductors and was a prime motivation for us to make a more systematic study of the influence of the inhomogeneities that are present in wires. We believe that the broad range of performance boundaries for wires in the literature is mostly a side effect of property distributions caused by these inhomogeneities, combined with the use of either arbitrary or variable criteria for $\mu_0 H_{c2}$.

The general consensus is that bronze-route wires are strongly inhomogeneous with respect to Sn content and that the gradients are on the order of 1-5 at. % Sn/μm[6,11]. Analysis of inhomogeneities in bronze-route wires are, however, complicated by their very small A15 layer thickness on the order of 1-2 μm, imperfect circular symmetry and the fact that the higher $T_c$ A15 layers surround the lower $T_c$ regions. The thin layers make accurate measurements of the Sn gradients by electron probe microanalysis impossible, imposing the more complex, more local and less accurate study by transmission electron microscopy. Analysis of the property distributions by magnetic characterizations are also not possible, since the outer higher $T_c$ regions magnetically shield the lower $T_c$ sections.

Table I: Overview of the investigated materials. Depicted are the sample names, a short identification number, the heat treatment for each sample, the wire type and the way each sample was mounted. The next column gives the additional elements that were present during formation of the A15 layers. In all wires, Cu was present close to the A15 formation area. The next column gives the non-Cu critical current density for each wire, either measured in transport or estimated from magnetization data. The last column gives the non-Cu current density in the samples during measurement of resistive transitions.

| Sample name | ID | Heat treatment | Type | Ø [mm] | Mounting procedure | Additions present | $J_{c\,non-Cu}$[b] 4.2K, 12T [A/mm$^2$] | Resistive data $J_{nonCu}$ [A/mm$^2$] |
|---|---|---|---|---|---|---|---|---|
| SMI ternary PIT[a] | B34-4h | 4h/675°C | PIT | 1.0 | GE Varnish on Cu | Cu, Ta | -- | 0.2 |
| SMI ternary PIT[a] | B34-16h | 16h/675°C | PIT | 1.0 | GE Varnish on Cu | Cu, Ta | 870[c] | 0.2 |
| SMI ternary PIT[a] | B34-64h | 64h/675°C | PIT | 1.0 | GE Varnish on Cu and Stycast on Ti-6Al-4V | Cu, Ta | 2250 | 0.2 |
| SMI ternary PIT[a] | B34-768h | 768h/675°C | PIT | 1.0 | GE Varnish on Cu | Cu, Ta | 2170[c] | 0.2 |
| SMI reinforced ternary PIT | B134 | 80h/675°C | PIT | 0.6 | Stycast on Ti-6Al-4V | Cu, Ta | 1961 | 0.6 |
| SMI binary PIT | B27 | 128h/675°C | PIT | 1.0 | Stycast on Ti-6Al-4V | Cu | 1955[c] | 0.2 |
| Vacuumschmelze ternary bronze | VAC | 220h/570°C+ 175h/650°C | Bronze | 0.8 | Stycast on Ti-6Al-4V | Cu, Ta | 556 | 0.5 |
| Furukawa ternary bronze[a] | FUR | 240h/650°C | Bronze | 0.8 | Stycast on Ti-6Al-4V | Cu, Ti | 582 | 0.5 |
| UW-ASC bulk[a] | Bulk | 320h/1020°C | Bulk, Sintered | 0.8×1.4 | GE Varnish on Ti-6Al-4V | -- | -- | 0.004 |

[a] Samples having a data inconsistency of ~4%.  [b] At a voltage criterion of 10µV/m.  [c] Estimated from magnetization data

Over the past decade, a Powder-in-Tube (PIT) process has become available as an alternative approach to Nb$_3$Sn wire manufacture. The filaments in modern PIT wires[24] are, in contrast to bronze conductors, nearly perfectly cylindrical, contain thick A15 layers on the order of 5-10 µm and have the higher $T_c$ A15 located on the inside, making them magnetically transparent. Initial compositional analysis indicated that they are much more homogeneous with respect to Sn content than bronze wires, exhibiting gradients on the order of 0.35 at. % Sn/µm[25, 26]. Magnetic (VSM) characterizations of these wires up to 14 T have indicated a large separation between $\mu_0 H_K(T)$ and $\mu_0 H_{c2}(T)$ which was correlated to Sn gradients. The latter were visualized by a reduction of the difference between $\mu_0 H_K$ and $\mu_0 H_{c2}$ with increasing reaction times, which are associated with a much stronger rise of $\mu_0 H_K(T)$ than $\mu_0 H_{c2}(T)$[27, 28]. These initial conclusions have recently been supported via $H$-$T$ characterizations on homogeneous bulk materials[29] as well as modeling[30].

A second important difference between wires and thin films and crystals is the presence of Cu and ternary additions as e.g. Ta and Ti. Cu is believed to lower the A15 formation temperature[31] and only to exist at the grain boundaries[32], but its influence on the critical properties is uncertain[29]. The general consensus on ternary additions as Ta and Ti is that they introduce scattering sites, thereby raising the resistivity and thus $\mu_0 H_{c2}(0)$[18, 33]. These ternary additions may also prevent the formation of tetragonal phases which reduce $\mu_0 H_{c2}(0)$[20, 34].

A third characteristic of wires is the presence of residual strain. Macroscopically, the strain dependence on $\mu_0 H_K(T)$ is well understood in terms of axial strain[35-37], or deviatoric strain[38, 39], but its influence on the complete $H$-$T$ phase transition is still uncertain[40]. The large Cu matrix enforces a compressive axial strain on the A15 layers due to the larger contraction of the Cu with respect to the A15. This causes a thermal pre-compression in the A15 layer which is spatially inhomogeneous[19].

The combined effects of Sn gradients, the presence of Cu in the A15 layer, the influence of ternary additions and the presence of strains make it questionable whether the $H$-$T$ knowledge, derived from uniform, well defined laboratory samples can directly be applied to wires. These considerations stimulated our systematic investigation of the effects of these factors on the $H$-$T$ phase transition in inhomogeneous Nb$_3$Sn wire systems.

A secondary advantage of a systematic characterization of the $H$-$T$ phase transition in wires is that it also yields a better understanding of the reasonable performance boundaries for the critical current density ($J_c$). Practical scaling relations for $J_c$[36, 37, 41-43] depend on the exact behavior of $\mu_0 H_{c2}(T)$, but are empirically based and generally rely on Kramer extrapolations of lower field data. Moreover, the zero temperature Kramer[14] extrapolated critical field ($\mu_0 H_K(0)$) often results in values far beyond 30 T[40]. This seems unrealistically high in comparison to actual measured results, which are limited to about 29.5 T for higher resistivity, slightly Sn-poor pure Nb-Sn film[2].

Expressions for the $H$-$T$ phase transition, derived directly from the BCS theory, are readily available in the literature[44-52], and have been applied with success to well defined laboratory specimens[1, 17, 53, 54], but most of the $H$-$T$ data on wires are empirically described[36, 37, 41]. This leads to continuing discussions in the literature, partly resulting from the use of extrapolated values of $\mu_0 H_{c2}(0)$ without taking account the inhomogeneous nature of the conductors. Although such empirical relations in principle yield sufficient accuracy for scaling of $J_c$[42], large discrepancies occur between the fitted values for $\mu_0 H_K(0)$ and $T_c(0)$ and actual measurements[40]. However well justified empirically, this discrepancy clearly needs understanding.

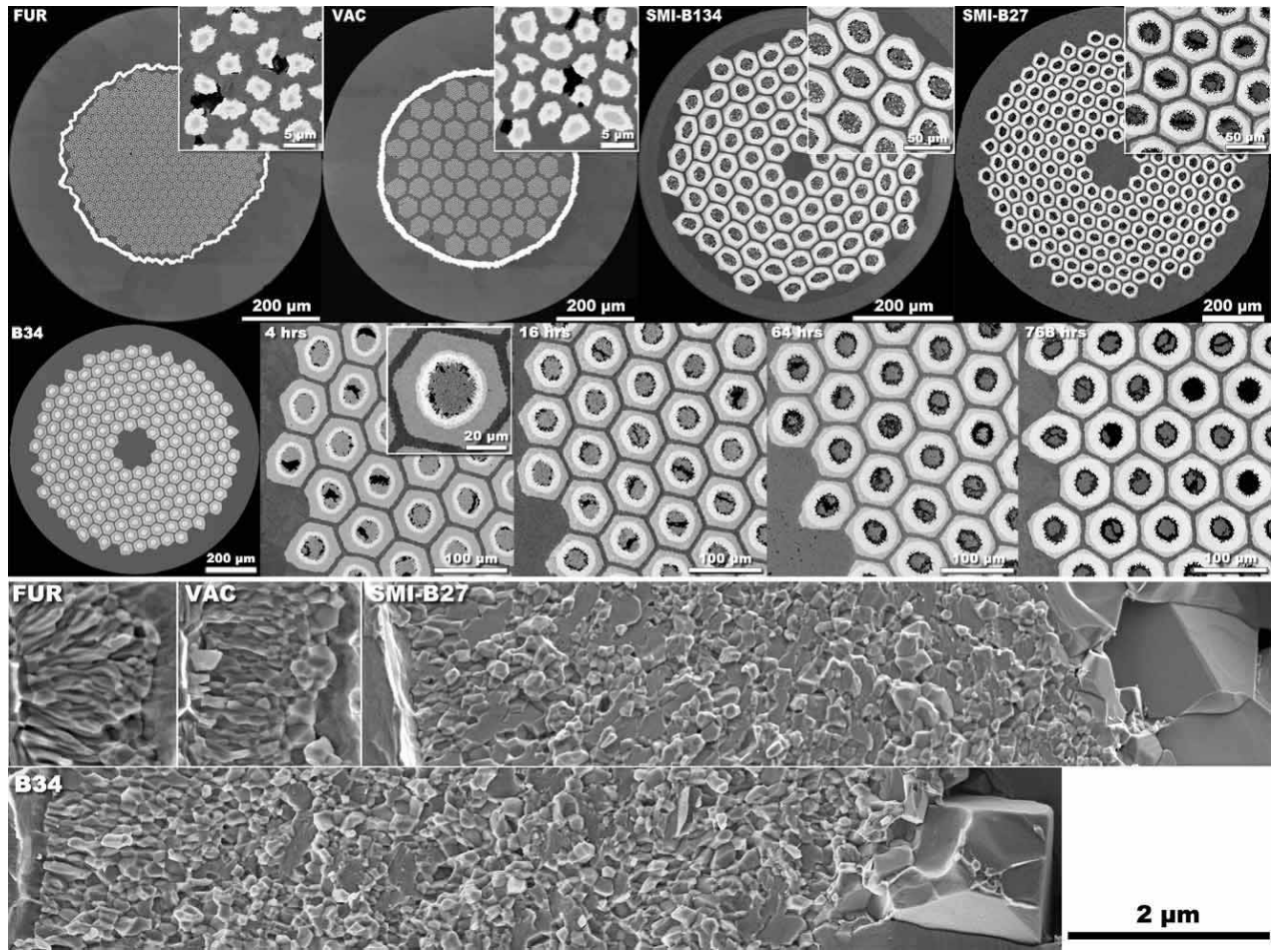

Figure 1: Field Emission Scanning Electron Microscope (FE-SEM) cross-section images of the investigated wires. Top section: Upper row from left to right: A ternary ITER type bronze process wire with Ta diffusion barrier manufactured by Furukawa (FUR) with a close-up of the filamentary region, a ternary ITER type bronze process wire with Ta diffusion barrier manufactured by Vacuumschmelze (VAC) with a close-up of the filamentary region, a ternary Powder In Tube (PIT) wire enclosed by a CuNb reinforcement tube manufactured by Shape Metal Innovation (SMI) plus close-up of the filamentary region and a binary PIT wire produced by SMI plus close-up of the filamentary area. Bottom row from left to right: An overall cross-section of a ternary PIT wire produced by SMI plus close-ups of the filamentary area for reactions of 4, 16, 64 and 768 hours at 675 °C respectively, indicating the progression of the A15 layer growth versus time. In the additional single filament close-up for 4 hours reaction an initial $Nb_6Sn_5$ phase is visible in between the core region and the formed $Nb_3Sn$. Bottom section: Close-ups on the A15 regions in the two bronze wires and in the binary and ternary (64 h-675 °C) PIT conductors. The Sn sources in these close-ups are located at the right side of the A15 layers. More detailed conductor specifications and heat treatment data can be found in Table I.

The goal of this paper is threefold. First we would like to understand the behavior of $\mu_0 H_{c2}(T)$ for inhomogeneous conductors independent of extrapolation method. Second, we would like to investigate the influences of compositional inhomogeneities, strain and the presence of additional elements on $\mu_0 H_{c2}(T)$. To achieve this, we have selected a sample set that contains variations in manufacturing technique, composition and strain state. Third, we would like to generalize the results by exploring whether an overall description for the $\mu_0 H_{c2}(T)$ behavior of $Nb_3Sn$ is possible, independent of measuring technique, criterion for the critical field, strain or compositional variation. We do this by investigating the validity of the Maki-DeGennes (MDG) equation, which is the simplest form of the available theoretical descriptions.

The paper is organized as follows. In Section II the sample materials are presented and in Section III the experimental procedure is explained. In Section IV the experimental data are presented. First we show that inhomogeneities can be visualized by electro-magnetic characterizations and we concentrate on the effects of inhomogeneity reduction by extended heat treatment times on the full *H-T* phase transitions. Second we make comparisons between various conductor types and then we investigate general aspects of the overall *H-T* data. In Section V we discuss the inhomogeneity effects on $\mu_0 H_{c2}(T)$, make comparisons between high field resistive and transport $J_c$ data and analyze the validity of the theoretical fits. Our overall conclusions are presented in Section VI.

**II. SELECTED SAMPLE MATERIALS**

In Table 1 an overview is presented of the selected samples with concise identification, the reaction conditions, fabrication procedure, sample cross-section, mounting technique, alloying additions, the non-copper $J_c$ and the non-copper resistive transition measuring current density. Cross-sections of the wires are shown in Figure 1.

The ITER type Furukawa (FUR) and Vacuumschmelze (VAC) wires were selected because they have been characterized extremely well in transport critical current density in various laboratories during the ITER benchmark tests[36,55,56]. They are representative of ternary conventional bronze-route manufacturing methods, have a relative low $J_c$, small filament size and probably large Sn gradients across the A15 areas. The Powder-In-Tube conductors from Shape Metal Innovation (SMI) were chosen since, due to their 'inside out' design, they are magnetically transparent[25,26], have a relatively large A15 cross section enabling microscopic composition analysis and are therefore favorable research wires that have already been extensively characterized[27,28]. They exhibit close to present record current densities and have, compared to bronze-route wires, less steep Sn gradients. A binary (B27) and two ternary versions (B34 and B134) of this wire type were selected. The main difference between the ternary versions is the presence of a Cu-Nb reinforcement tube around the outside of B134, as opposed to the pure Cu matrix used for the regular ternary (and binary) wire. The Cu-Nb reinforcement of B134 results in slightly more deformation of the outer filaments and a larger compressive A15 strain state. The pure Cu matrix ternary conductor (B34) was additionally selected for study of inhomogeneity effects using increased reaction times, which result in increased A15 layer thickness and less steep Sn gradients[28]. A bulk sample, produced at the University of Wisconsin's Applied Superconductivity Center (UW-ASC), was used to investigate the behavior of pure binary, Cu-free $Nb_3Sn$.

The first set of samples is a ternary PIT wire (B34) that was heat treated for various times at 675 °C. A reaction time of 64 hours was recommended by the manufacturer (SMI) to yield optimal $J_c$. A specific mounting procedure was applied in an attempt to force the samples into a reproducible, although unknown, strain state after cool down to test temperatures. Four wires were mounted with General Electric Varnish onto a copper substrate. One sample, reacted for 64 hours was mounted with Stycast 2850FT on a Ti-6Al-4V substrate of lower thermal contraction than Cu (~0.18 % versus ~0.35 %). Standard $J_c$ barrels were made from Ti-6Al-4V and Stycast was applied to mount the wire rigidly on the barrel as in the ITER benchmark measurements[42]. Inter-comparisons of short sample resistive transitions (as used in this paper) and standard $J_c$ data require the same mounting technique in an attempt to create identical strain states after cool down.

An overview of the ternary PIT wire (B34) and a magnified view of the filaments for reactions varying from 4 to 768 hours at 675 °C is shown in the second row of Figure 1. The wire consists of hexagonal Nb-7.5 w. % Ta tubes, embedded in a pure Cu matrix. The tubes are filled with powder, consisting of a Sn rich Nb-Sn intermetallic combined with Sn and Cu which acts as the Sn source for the solid-state diffusion reaction. To the right of the overall cross-sections are enlarged filament sections for 4, 16, 64 and 768 hour reactions at 675 °C. The A15 formation (light gray) progresses with increased reaction time. The inset for the 4 hour cross-section shows two phases that are present at the core-A15 interface in the initial stage of the reaction. The light gray region around the core is $Nb_6Sn_5$ and the darker gray shell around this is A15. At 16 hours the $Nb_6Sn_5$ has disappeared. The bottom row of Figure 1 shows a magnified view across the whole A15 layer in the ternary PIT wire reacted for 64 hours at 675 °C. The core is located to the right and the Nb(Ta) tube to the left. Large, 1-2 µm diameter A15 grains arising from conversion of the initial $Nb_6Sn_5$ are found next to the core and are not believed to contribute to $J_c$. The large central region is fine-grain A15 with an average grain size of about 140 nm. Similar thick, fine-grain A15 layers have been observed in high $J_c$ Oxford Instruments Superconducting Technology Internal Sn strand material[57], which also has an abundant Sn source in the core. Columnar grains are found in the last ~1 µm before the Nb(Ta) tube. A possible cause of the columnar grains is their low Sn content. Thermodynamics dictates that a portion of the A15 layer must be Sn-poor. We also noted that the 768 hour reaction extent was not much greater than after the 64 hour heat treatment, suggesting that the reaction quickly tends to exhaustion.

The second sample type in Table 1 is also a ternary PIT wire manufactured by SMI (B134) which is similar in build to B34 but in this case with a Cu-Nb reinforcement tube surrounding the Cu matrix. It can be expected that the reinforcement results in a larger thermal pre-compression in the A15 from reaction temperature down to room temperature caused by the higher strength of the Cu-Nb as compared to a pure Cu matrix. The third cross-section in the top row of Figure 1 shows that the Cu-Nb reinforcement results in additional deformation of the outer filaments as compared to the not-reinforced PIT wires.

The next sample in Table 1 is a binary PIT wire (B27), again manufactured by SMI. The only difference from the ternary PIT (B34) wire according to the manufacturer is the use of pure Nb tubes. This wire was heat treated beyond the manufacturer's recommended time which was 47 hours at 675 °C since earlier magnetic characterizations on this wire indicated that a longer reaction of 128 hours yields slightly improved properties[28]. The top right picture in Figure 1 shows the overall cross-section, while the A15 layer is shown in the lower part for B27, again with the Sn source located at the right side. Large grain A15 is visible at the core interface as in the ternary version of this wire. The remainder of the A15 layer is comparable to the central region of the ternary wire, i.e. fine grain (~140 nm) A15 but the columnar grains at the Nb-tube/A15 interface are less pronounced.

The Vacuumschmelze wire was heat treated with the manufacturer's recommended reaction. This bronze-route wire uses Nb-7.5 w. % Ta rods that are embedded in a Sn rich bronze matrix that is in its turn surrounded by a Ta diffusion barrier and embedded in a Cu matrix. It is visible in the second cross-section from Figure 1 that the Nb rods are bundled into sub-elements. In the inset ~5 µm filaments are shown with the light gray areas representing the A15. The voids might be caused through Sn depletions in the bronze or possible polishing side effects. The A15 layer cross-section in the bottom part of Figure 1, which has the bronze located to the right, shows that columnar grains are formed over a large area on the A15/Nb(Ta) interface side. A much smaller fine grain equiaxed (high Sn) region which is only about 2-3 grains thick is visible at the bronze interface. A much lower Sn diffusion rate is characteristic of bronze wires as compared to the PIT wires.

The Furukawa bronze-route wire was also heat treated with the manufacturer's recommended reaction. This wire is constructed with Nb rods embedded in a Ti-containing Sn-rich bronze. The bronze is surrounded with a Ta diffusion barrier outside of which is pure Cu. From the upper left cross-section in Figure 1 it can be seen that the single stack of filaments are more evenly distributed than in the Vacuumschmelze sub-bundles. Voids are present in the bronze after reaction. The upper left picture in the bottom part of Figure 1 shows a ~1.5 μm thick A15 layer consisting mainly of columnar grains and an equiaxed layer that appears to be only about one-grain thick at the A15/bronze interface located at the right side.

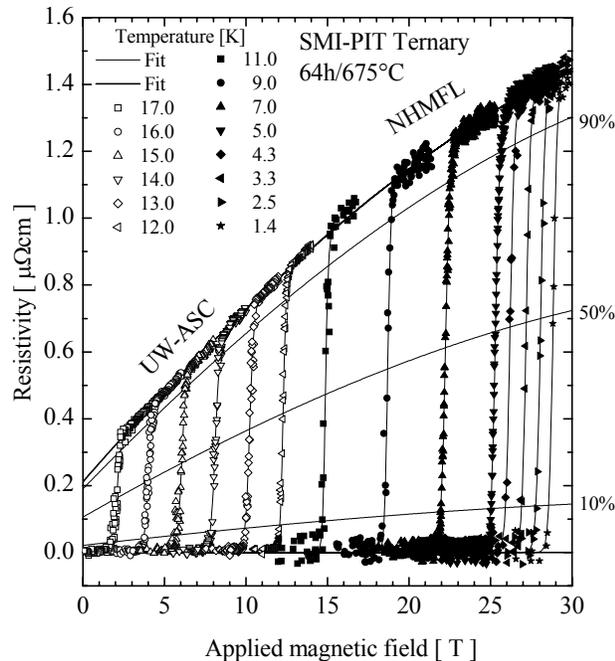

Figure 2: Representative resistive transitions for the ternary PIT wire (B34) as measured at the UW-ASC (open symbols) and the NMHFL (closed symbols). The upper envelope line represents an overall polynomial fit of the magneto-resistance and thin lines represent exponential fits to individual transitions using equation 1 to overcome the noise that is present in the high field data. Intersections can be made at various percentages of the normal state resistivity, as indicated by the 10 %, 50 % and 90 % lines to arrive at different criteria for $\mu_0 H_{c2}$.

The nominally 24.4 at. % Sn bulk sample was cut into a needle for resistive characterization and mounted with GE Varnish on Cu. It was not fully equilibrated and exhibits therefore three-dimensional inhomogeneities in contrast to the wire samples, which show mainly radial Sn gradients. Cross-sections of this not equilibrated sample show local A15 areas that are stoichiometric, as well as others low in Sn[29]. This inhomogeneous A15 distribution results in the measuring current sampling a greater range of A15 Sn compositions than in the wires which are longitudinally rather homogeneous, although radially inhomogeneous. The sample is a cutout from a larger bulk section, produced by Hot Isostatic Pressing (HIP). Field Emission Scanning Electron Microscopy (FESEM) analysis, more details on the production, as well as more extensive characterization on a larger bulk sample set, were published earlier[29].

### III. EXPERIMENTAL PROCEDURE

#### A. Resistive characterizations

The majority of the *H-T* investigations in this paper were performed by resistive characterizations at constant temperature and ramping field. Data for fields ranging from 12-30 T were taken at the National High Magnetic Field Laboratory (NHMFL), Tallahassee FL using a resistive high field magnet while data up to 15 T were obtained using a superconducting solenoid at the UW-ASC. The resistive technique was chosen since this method is easy to set up at the NHMFL and is consistent with our earlier transport $J_c$ characterizations. A disadvantage is that only part of the full property distributions is detected in this way, due to relatively low excitation current densities. A second disadvantage is that the normal state signal of the wires typically was only on the order of 200 nV at zero field, whereas the background noise in the high field resistive magnet area was typically on the order of 50-150 nV. The strong magneto-resistance of the Cu matrix of the wires fortunately delivered an enhancement factor of 4-10 in signal at high field. A nice advantage of the resistive technique is that it is possible to characterize multiple samples at once, saving valuable high field magnet time. The wire sections in the small current resistive measurements were about 8mm long with a voltage tap separation of 1-3 mm and current taps at both ends. The dimensions of the bulk needle sample were $7.75 \times 1.42 \times 0.80$ mm$^3$. Samples were mounted on a strain-fixing substrate, according to Table I. The resistive measurements were performed at a small current (generally 0.2-0.6 A/mm$^2$, see Table I), constant temperature and a swept field $d\mu_0 H/dt = 10$ mT/s. Typical variations in temperature during a constant temperature sweep ranged from ± 5 mK in the best case to ± 50 mK in the worst case. Low current, constant field, swept temperature ($dT/dt = 3$ mK/s) measurements were confirmed to deliver the same $\mu_0 H_{c2}(T)$ data as the swept field characterizations. A detailed description of the technique has been published elsewhere[40].

A systematic inconsistency of ~4 % exists between the NHMFL and UW-ASC datasets for the variable reaction ternary wires (B34), the Furukawa wire and the bulk sample, as indicated in Table I. Temperature measurements in both institutes were confirmed to be correct and the same probe is used in both systems. The magnets in both institutes were recalibrated but no errors were detected. The Furukawa sample has been re-measured in a different resistive magnet over the full temperature range. These results average out the jump that was initially visible between the separate low- and high field datasets. An error source could also be that the strain changes with thermal cycling, despite the attempts to force the samples into a reproducible strain state. An axial strain difference of ~0.06 % could result in the observed ~4 % inconsistency[36]. No corrections are made on the data since no unambiguous error source could be found and the data should therefore be regarded to be reliable within ~± 2 %.

#### B. Data reduction techniques

A typical set of resistively measured transitions is shown in Figure 2 for a ternary PIT wire (B34) reacted for 64 hours at 675 °C. The measured transitions are fitted by a shifted and normalized tangent hyperbolic function multiplied by a polynomial term which describes the magneto-resistance[40]:

$$\rho(\mu_0 H) = \frac{\exp\left[4e\left(\frac{\mu_0 H - \mu_0 H_{1/2}}{\mu_0 H_W}\right)\right]}{\exp\left[4e\left(\frac{\mu_0 H - \mu_0 H_{1/2}}{\mu_0 H_W}\right)\right]+1} \times \\ \times \left(\frac{\rho(0) + C_1 \mu_0 H + C_2 (\mu_0 H)^2}{1 + C_3 \mu_0 H}\right) \quad . \quad (1)$$

where $C_{1-3}$ are constants and $\mu_0 H_{1/2}$ and $\mu_0 H_W$ represent the applied magnetic field value at half the transition height and the transition width respectively. This equation yields excellent fits to the data as can be seen in Figure 2. The fit works as a low pass filter, especially for the high field noise. Critical fields at various criteria (e.g. 10 %, 50 %, 90 % normal state) can easily be obtained from the fits.

The excitation current dependence of the width of transitions was investigated with respect to $\rho(T)|_{H=0}$ transitions in an earlier publication[40]. It was found that the width of the $\rho(T)|_{H=0}$ transition broadens with increased excitation current, but that the upper shelf (i.e. above ~90 % normal state) is independent of excitation current density. It is assumed that this observation holds for the present $\rho(\mu_0 H)|_T$ characterizations, since a higher excitation current density results in the detection of a larger property distribution as more of the lower $\mu_0 H_{c2}$ A15 is probed. This means that a resistive transition only samples a small fraction of the A15. The majority of the detected property distributions lie between 10 % and 90 % normal state, as indicated in Figure 2. The upper shelves of the transitions (i.e. 90-99 %) are interesting since they represent the highest $\mu_0 H_{c2}$ A15 sections that are detected in the samples. Magnetic characterizations on similar samples have indicated that magnetically determined $\mu_0 H_{c2}(T)$ using the onset of superconducting behavior as criterion, approximately correspond to the onset of superconductivity (i.e. a 99 % normal state criterion) in the resistive transitions[28, 58]. The low noise threshold of our superconducting-magnet measurements (up to 15 T) allows for the use of any transition criterion but the higher noise levels of the Bitter magnets used at high fields call into question the validity of transition criteria above 90 %. A criterion of 99 % at high field can however be found by extrapolation of the resistivity fit of the unambiguous data from ~10 % to ~90 % $\rho_{Tc}$. Assuming similar transition widths in lower- and higher field enabled data for the full $\mu_0 H_{c2}(T)$ transition to be extracted.

An overall fit of the resulting $H$-$T$ data sets was used to extract $\mu_0 H_{c2}(0)$ and $T_c(0)$ from the data points at finite temperature and field. The measured $\mu_0 H_{c2}(T)$ phase boundaries were least squares fitted with the Maki-DeGennes[44-47] (MDG) description[40]:

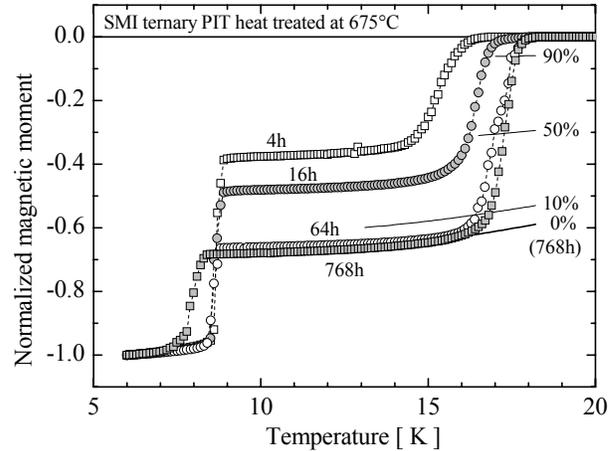

Figure 3: SQUID magnetometer data on ternary PIT (B34) wires after 4, 16, 64 and 768 hours at 675 °C, taken by Fischer[28]. The magnetic moment versus temperature was obtained through zero field cooling of the samples to 5 K, application of a 5 mT field parallel to the wire axis to introduce shielding currents and raising the temperature while registering the magnetic moment. The lines depict intersections at 10 %, 50 % and 90 % normal state for the 768 hour sample where 0 % is defined at the lower end of the A15 transition, depicted by the bold line. These lines represent criteria that can be used to derive $T_c$(5m T).

$$\ln\left(\frac{T}{T_c(0)}\right) = \psi\left(\frac{1}{2}\right) - \psi\left(\frac{1}{2} + \frac{\hbar D \mu_0 H_{c2}(T)}{2\phi_0 k_B T}\right). \quad (2)$$

The function uses only two fitting parameters, namely $T_c(0)$ and the diffusion constant of the normal conducting electrons $D$. The other parameters are the reduced Planck constant ($\hbar$), the magnetic flux quantum ($\phi_0$) and the Boltzmann constant ($k_B$). Although fuller characterization of paramagnetic limitation and spin-orbit scattering effects, as well as strong coupling corrections and non-spherical Fermi surfaces is provided by Werthamer, Helfand and Hohenberg[48-52], we did not find that this added any additional insight into our data on inhomogeneous Nb$_3$Sn conductors.

## IV. RESULTS

### A. Magnetic tests of inhomogeneity in variably reacted wire

The ternary PIT wires (B34) with 4, 16, 64 and 768 hour reactions at 675 °C were investigated by SQUID magnetization measurements[28]. The resulting normalized magnetic moments as function of temperature at an applied field of 5 mT are shown in Figure 3. At 5 K full flux exclusion can be observed.

Table II: Summarized zero temperature upper critical field and zero field critical temperature data.

| Criterion | Parameter | B34-4h | B34-16h | B34-64h | B34-768h | B134 | FUR | VAC | B27 | Bulk |
|---|---|---|---|---|---|---|---|---|---|---|
| 10% | $\mu_0 H_{c2}$ [T] | 27.0 | 27.4 | 28.9 | 29.0 | 28.2 | 28.0 | 27.3 | 26.6 | 26.5 |
|  | $T_c$ [K] | 16.8 | 17.1 | 17.6 | 17.4 | 17.8 | 17.2 | 17.4 | 17.8 | 16.5 |
| 90% | $\mu_0 H_{c2}$ [T] | 28.1 | 28.4 | 29.4 | 29.5 | 29.1 | 28.9 | 28.5 | 27.4 | 28.3 |
|  | $T_c$ [K] | 17.2 | 17.3 | 17.8 | 17.7 | 17.9 | 17.4 | 17.7 | 17.9 | 16.6 |
| 99% | $\mu_0 H_{c2}$ [T] | 28.8 | 29.0 | 29.7 | 29.7 | 29.6 | 29.3 | 29.2 | 27.8 | 29.3 |
|  | $T_c$ [K] | 17.3 | 17.5 | 17.9 | 17.8 | 18.0 | 17.5 | 17.8 | 17.9 | 16.7 |

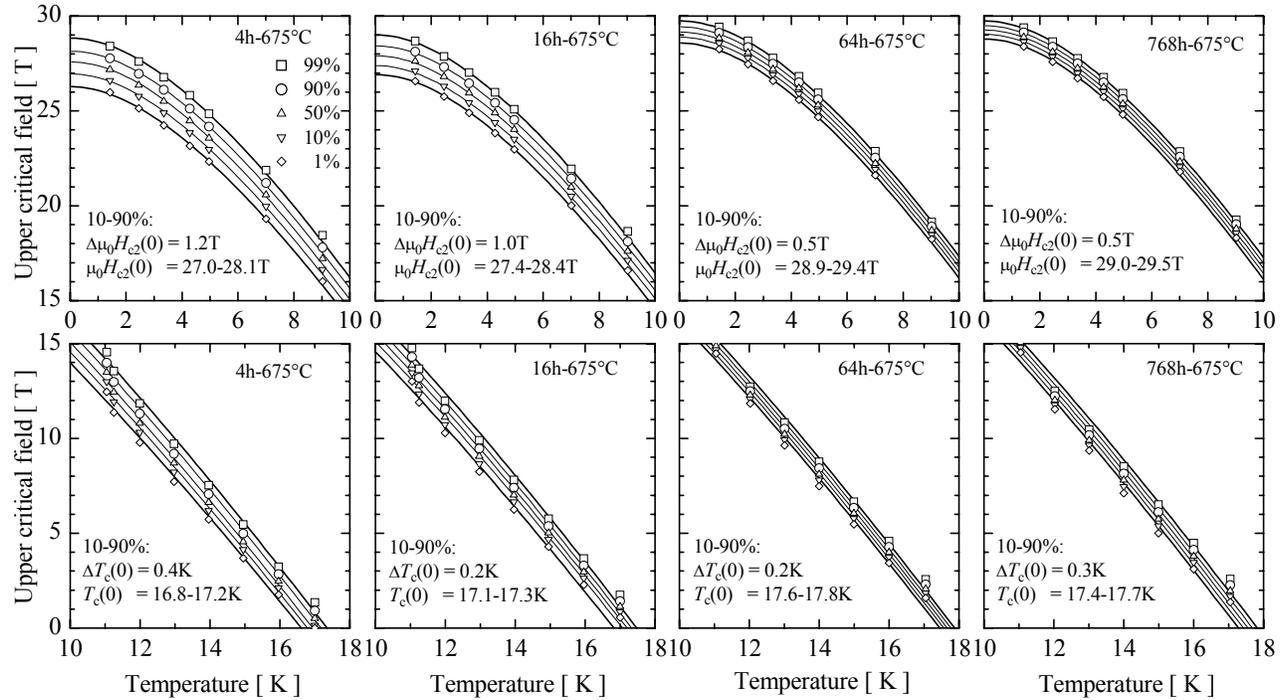

Figure 4: Reduction of the property gradients via extended heat treatment visualized via resistive *H-T* data on SMI ternary (B34) wire. The points in these plots were derived from the resistivity fits on the data and the lines were calculated with the MDG description. The 1 % and 99 % lines are used to quantify the extremes of the measured transitions. The main part of the detected property distribution occurs between 10 % and 90 %. Clearly visible is the reduction of the $\mu_0 H_{c2}(0)$ transition width (10-90 %) with increasing heat treatment time from 1.2 T to 0.5 T. The best A15 sections that are detected ($\mu_0 H_{c2}(0)$-99 %) only increase by ~3 %, but the lower end of the detected property distribution (($\mu_0 H_{c2}(0)$-1 %)) increases by ~10 %. The reduction of detected $T_c(0)$ width, as derived from the MDG fits is less obvious but still present, resulting in a width (10-90 %) reduction from 0.4 K to 0.3 K.

Around 8.5 K the initial transition occurs for the 4 h, 16 h and 64 h reactions. The first transition in the 768 h heat treatment lies substantially lower, starting at 7 K and ending at 8.1 K. The main A15 transitions occur between 14 and 18 K. The steady progress of the reaction is clearly visible. The increased signal amplitude for longer time indicates an increase in A15 quantity, in agreement with the SEM observations. The Sn enrichment (or A15 quality improvement) is visualized by the shift of the transitions to higher temperatures, combined with a reduction of the transition widths. It is notable that extension of the reaction from 64 to 768 hours slightly increases the upper $T_c$, while barely enhancing the signal magnitude at $T = 10$ K. However, a depression of the "Nb" transition to ~7.5 K indicates some penetration of the A15 layer through the Nb which makes low $T_c$ A15 phases visible.

The lines at 0, 10, 50 and 90 % in Figure 3 represent intersections that can be made on the 768 hour A15 transition. The bold line is an extrapolation of a least squares fit on the data points between 10 and 14 K defining a 0 % normal state for the A15 transitions. Intersections were derived in a similar way for the 4, 6 and 64 hour reactions. The 10 % and 90 % normal state intersections on the A15 transitions above 10K were used to define $T_c$-10 % (5 mT) and $T_c$-90 % (5 mT) for plotting in Figure 5.

**B. Resistive visualization of inhomogeneity reduction via elongated reactions**

To expand the inhomogeneity investigations on the ternary PIT (B34) wire to the whole *H-T* range, we switched to the resistive characterization of the $\mu_0 H_{c2}$ transition, as shown in Figure 4 and summarized in Table II, where the points represent the 1, 10, 50, 90 and 99 % normal state-resistance fits to equation 1. The lines are MDG fits to equation 2, using $T_c(0)$ and $D$ as free parameters. The majority (~50 %) of the detected property distributions are present between the 10 and 90 % normal state points. The UW-NHMFL data inconsistency mentioned earlier is visible around $T = 11$-$12$ K, causing deviations from the overall $\mu_0 H_{c2}(T)$ fits at $T = 12$-$17$ K.

It can be seen that the $\mu_0 H_{c2}(0)$ transition width reduces from 1.2 T to 0.5 T as the reaction increases from 4 to 768 hours at 675 °C. The better A15 sections ($\mu_0 H_{c2}(90\%)$) increase by 1.4 T from 28.1 T to 29.5 T. The lesser A15 sections ($\mu_0 H_{c2}(10\%)$) in contrast increase by 2.0 T from 27.0 to 29.0 T and their stronger rise in comparison with $\mu_0 H_{c2}(90\%)$ is the origin of the transition width reduction with increasing reaction time. The majority of the reduction occurs between 16 h and 64 h. The width reduction effect on $T_c(0)$ is less pronounced but still visible, changing from 0.4 K to 0.2 K minimum (10-90 %). The best A15 sections detected (99 %) rise by ~3 % in $\mu_0 H_{c2}(0)$ from 4 to 768 hours, whereas the lower transition (1 %) rises ~10 %. The extremes in $T_c(0)$ change by ~3 % for the best detected sections and ~4 % for the lower transition. Optimum critical current densities, expressed as the critical current divided by the total Nb(Ta)+A15+core package area, are achieved at 64 hours. Maximum $J_c$ appears to correlate to the rise to maximum $\mu_0 H_{c2}$ seen on going from 16 to 64 hours, before any Sn escapes from the Nb(Ta) tube or additional grain growth that lowers the flux pinning occurs.

The MDG fits at 12 K derived from the resistive transition data are compared to the magnetic data in Figure 5. The points and dotted lines are derived from the resistive data in Figure 4 and the shaded areas are the magnetically (VSM and SQUID) derived property distributions from Figure 3 and a reproduction from earlier published[27] VSM data at $T$ = 12 K. The decreasing difference from ~5.4 T to ~2.4 T between the magnetically derived $\mu_0 H_K$(12 K) and $\mu_0 H_{c2}$(12 K) with increasing reaction time is a direct result of the Sn gradient reduction in the A15 layer, in agreement with recent simulations[30]. The reduction of the transition width in the resistive characterizations is less obvious (from ~1.1 to 0.6 T for the 10 and 90 % normal state resistance points) but still clear. The highest $\mu_0 H_{c2}$(12 K) as detected in the resistive characterizations (at 99 % normal state resistance) is identical to and increases in the same way as the VSM determined $\mu_0 H_{c2}$(12 K). The SQUID derived $T_c$(0) transition width reduces from 1.3 to 0.9 K, whereas the resistive transition width (between 10 and 99 % normal state resistance) reduces from 0.4 to 0.3 K. The SQUID derived $T_c$(0) values, however, are substantially lower than in the resistive data. The magnetic and resistive characterization techniques are therefore, apart from the highest detected $\mu_0 H_{c2}$ values, only qualitatively comparable. This stems from the fact that small current density resistive transitions sample only a small fraction of the A15 with the highest property.

### C. Inhomogeneity differences between the conductors

An overview of the collected $\mu_0 H_{c2}(T)$ data for all resistively characterized samples, reacted 'normally' (see Table I), is given in Figure 6 and summarized in Table II. The plots are again separated in temperature for enhanced visibility and developed as for the ternary PIT data in Figure 4.

The order from top to bottom in Figure 6 is approximately of increasing inhomogeneity. The transition widths with respect to $\mu_0 H_{c2}$(0) range from 0.5 T for the ternary PIT wire (B34) reacted at 64 hours, to 1.8 T for the bulk sample. The fitted $T_c$(0) variation ranges from 0.1 K for the binary and reinforced ternary PIT wires and the bulk sample to 0.3 K for the Vacuumschmelze bronze wire. It is interesting to observe the substantial differences that occur between the two bronze conductors. The Ti-alloyed Furukawa wire exhibits a smaller distribution (0.9 T and 0.2 K) and better overall properties than the Ta-alloyed Vacuumschmelze wire (1.2 T and 0.3 K).

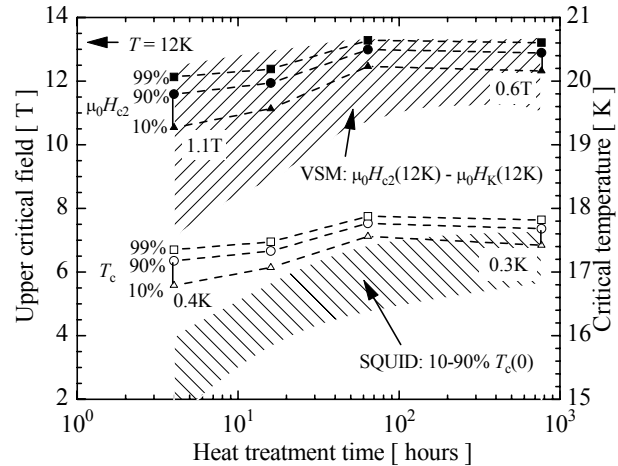

Figure 5: Summarized result from Figure 4 depicting the effect of reaction time on the resistively measured $\mu_0 H_{c2}$-99 %(12 K), $\mu_0 H_{c2}$-90 %(12 K), $\mu_0 H_{c2}$-10 %(12 K), $T_c$-99 %(0), $T_c$-90 %(0) and $T_c$-10 %(0) calculated from the MDG fits (dotted lines plus data points) in comparison with the SQUID and VSM data from Figure 3 and Fischer *et al.*[27] (depicted by the shaded regions). The general trends for the critical fields versus reaction time as measured resistively are similar to the magnetic data but the detected transition widths are much smaller in the resistive characterizations. The highest $\mu_0 H_{c2}$(12 K) values (i.e. at 99 % normal state resistance) for the resistive data coincide with the VSM determined $\mu_0 H_{c2}$(12 K) values. A difference is visible between the detected $T_c$(0) values in addition to the reduced transition width in the resistive characterizations.

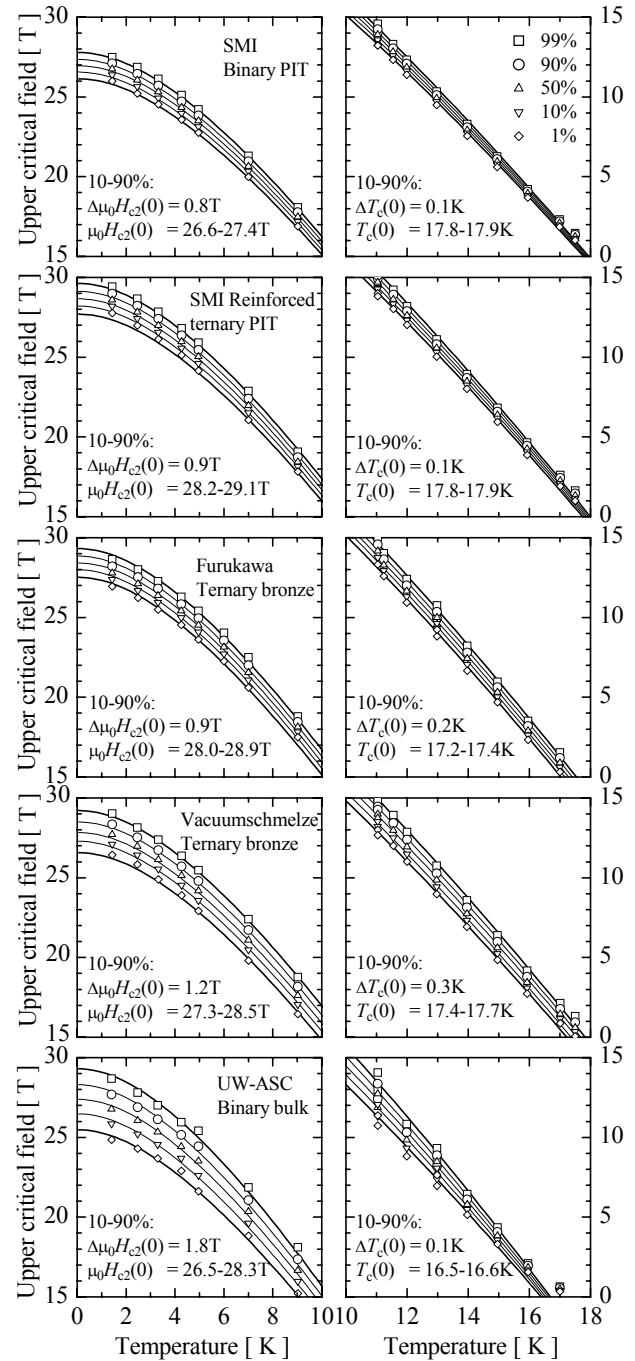

Figure 6: Collected $\mu_0 H_{c2}(T)$ data for resistively characterized samples from 1 to 99 % normal state. The plots are separated in temperature for enhanced visibility. Included are the range of values and detected property distribution widths for the 10-90 % normal state criteria. The points were derived from the resistivity fits on the data and the lines were calculated with MDG. The main part of the detected property distribution occurs between 10 % and 90 %. The 1 % and 99 % lines are used to quantify the extremes of the measured transitions. The order from top to bottom is approximately of increasing inhomogeneity.

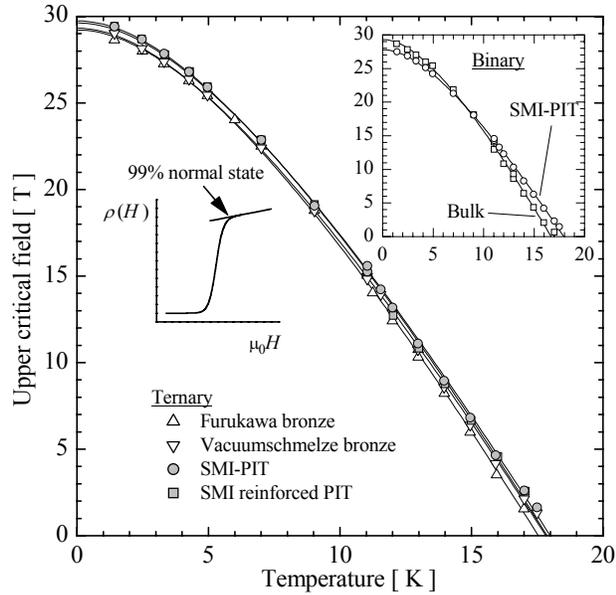

Figure 7: Comparison of the best A15 sections that were detected in the ternary wires. The inset shows the binary PIT wire plus the binary bulk needle sample. The points were derived from the resistivity fits on the data and the lines were calculated with the MDG description. It can be seen that the best A15 regions are present in all ternary wires and very comparable, i.e. $\mu_0 H_{c2}(0) = 29.5 \pm 0.3$ T and $T_c(0) = 17.8 \pm 0.3$ K. The binary PIT in comparison shows a reduced maximum $\mu_0 H_{c2}(0)$ (27.8 T) but a comparable $T_c(0)$ (18.0 K). The bulk needle has a reduced fitted $T_c(0)$ (16.7 K) but, in comparison to the binary PIT wire, a very high $\mu_0 H_{c2}(0)$ which is in the range of the ternary wires (29.3 T).

The bulk needle sample shows a very large distribution in $\mu_0 H_{c2}(0)$ of 1.8 T, or even 3.9 T if the outer extremes (1 and 99 %) are taken into account. This large spread is indicative of its large Sn inhomogeneity. This sample exhibits a three dimensional distribution of local areas of high and low Sn content[29] and no continuous high A15 quality current path therefore exists. The wire samples, in contrast, exhibit a radially distributed Sn gradient and thus have longitudinal parallel paths of different, but approximately constant Sn content. The ternary PIT wire (B34) is known to exhibit relatively low Sn gradients[25].

### D. The best properties detected in various conductors

The highest accessible $\mu_0 H_{c2}(T)$ data (at 99 % normal state resistance) for all samples are collected in Figure 7 and Table II. Considering the varying transition breadths in Figure 6, the highest $\mu_0 H_{c2}(T)$ A15 sections in the ternary wires are strikingly constant. These high $\mu_0 H_{c2}(T)$ A15 regions are present in all ternary wires and are very comparable: $\mu_0 H_{c2}(0) = 29.5 \pm 0.3$ T and $T_c(0) = 17.8 \pm 0.3$ K. A small difference can be observed between the bronze and PIT conductors, the highest detected phase boundaries in the bronze wires being slightly lower than in the PIT wires. This small difference in $\mu_0 H_{c2}(T)$ is in strong contrast to their large critical current density differences which are approximately 4 times higher in the PIT wires (see Table I).

The inset in Figure 7 compares the binary PIT wire to the binary bulk needle. The binary PIT shows a reduced maximum $\mu_0 H_{c2}(0)$-99 % of 27.8 T but a comparable $T_c(0)$-99 % of 18.0 K. The bulk needle has a reduced $T_c(0)$-99 % of 16.7 K but, in comparison to the binary PIT wire, a very high $\mu_0 H_{c2}(0)$-99 % of 29.3 T which is in the range of the ternary wires.

## V. DISCUSSION

### A. Overall behavior of the *H-T* phase boundary

The upper (90-99 %) transitions of all ternary wires are remarkably similar (Figure 7), in contrast to the large range of transitions widths (Figure 4 and 6) and grain morphology differences (Figure 1). The binary PIT wire has a suppressed $\mu_0 H_{c2}(0)$ while retaining a comparably high $T_c(0)$ in comparison to the ternary wires. The bulk needle exhibits a $\mu_0 H_{c2}(0)$ which is comparable to the ternary wires but shows a suppressed $T_c(0)$. Orlando et al.[2] have shown similar behavior in thin films with varying resistivity: Increasing the resistivity caused $\mu_0 H_{c2}(0)$ to rise, at the cost of a reduction in $T_c(0)$. Their optimal dirty film ($\rho_{Tc} = 35$ µΩcm) had $\mu_0 H_{c2}(0) = 29.5$ T and $T_c(0) = 16.0$ K at a 50 % resistive criterion, compared to 26.3 T and 17.4 K for a $\rho_{Tc} = 9$ µΩcm film. Our bulk needle exhibited a resistivity just above $T_c(0)$ of 22 µΩcm and $\mu_0 H_{c2}(0)$ and $T_c(0)$ at a 50 % normal state resistance criterion are 27.4 T and 16.5 K respectively. However, our bulk needle was far from fully homogenized, meaning that for a more honest comparison a higher resistive criterion might be more suitable. A 90 % normal state resistance results in $\mu_0 H_{c2}(0) = 28.3$ T and $T_c(0) = 16.6$ K. Both the 50 % and 90 % values appear consistent with the thin film data from Orlando et al., i.e. at $\rho_{Tc} = 22$ µΩcm our bulk needle phase boundary is positioned between the 9 and 35 µΩcm thin film data. In addition, a (partial) transformation to the tetragonal phase could also reduce $\mu_0 H_{c2}(0)$.

The upper (90-99 %) resistively determined transitions for the ternary PIT wire are identical to the magnetically (VSM) derived $\mu_0 H_{c2}$ data for the limited overlap that is available at $T = 12$ K. This comparison validates our conclusion that the highest detectable $\mu_0 H_{c2}$ can be probed using completely different characterization methods.

### B. Comparisons of resistive *H-T* data and transport $J_c$ data

To investigate how $\rho(\mu_0 H)|_{T=4.2\,K}$ data correlates to $J_c(\mu_0 H)|_{T=4.2\,K}$ characterizations, a Kramer plot of high field $J_c$ data on a ternary PIT wire (B34-64h) is combined with a resistive $\mu_0 H_{c2}(4.2\,K)$ transition on the same wire in Figure 8. The $J_c$ sample was mounted with Stycast on a helical Ti-6Al-4V barrel[36] with reduced diameter and the resistive $\mu_0 H_{c2}$ transition sample was mounted in an identical way to reproduce the same strain state. The Kramer plot of the $J_c$ data is indeed highly linear, almost up to $\mu_0 H_K(4.2\,K)$. The extrapolated value for $\mu_0 H_K(4.2\,K)$ (24.9 T) is in agreement with magnetic (VSM) Kramer data at 4.2 K found by Fischer[28] which were measured in the range $\mu_0 H = 6$-14 T, showed perfect Kramer linearity and also extrapolated to $\mu_0 H_K(4.2\,K) = 25$ T. This indicates a straight Kramer plot from $\mu_0 H = 6$ to ~23 T, the magnetic field value for which the critical current data starts to deviate from linear Kramer behavior. The observed linearity and small tail for the PIT wire is in agreement with recent modeling using actual measured Sn gradients as input data[30].

Two particular points of the resistive transition are important: One occurs at 10 μV/m, the voltage criterion used in the transport $J_c$ characterization, and one at 99 % of the normal state resistance, identifying the highest observed value for $\mu_0 H_{c2}(4.2\ K)$. The 10 μV/m point on the resistive transition represents a critical current density of 0.2 A/mm² at 4.2 K and ~25 T (the excitation current) and yields $0.2^{1/2} 25^{1/4} = 1\ A^{1/2} T^{1/4}$ on the Kramer axis. The transport $J_c$ data at 4.2 K and 25 T yield $3\ A^{1/2} T^{1/4}$, which is a very reasonable agreement.

The linear Kramer extrapolation of the measured $J_c$ data yields $\mu_0 H_K(4.2\ K) = 24.9$ T as the scaling critical field for $J_c(4.2\ K)$. The highest $\mu_0 H_{c2}(4.2\ K)$ detected in the resistive transitions, yields 26.9 T for 99 % normal state resistivity. We can now estimate the hypothetical $J_c(4.2\ K)$ gain which would be achieved if the entire A15 layer would be of the highest $\mu_0 H_{c2}(4.2\ K)$ quality and grain size and grain boundary densities could be retained, in a way similar as recently done by Cooley et al.[30] who used an identical approach based on modeling the A15 Sn gradients. We use the definition of the Lorentz force that balances the bulk pinning force:

$$J_c(\mu_0 H) \times \mu_0 H = -F_P(\mu_0 H), \qquad (3)$$

and combine this with the temperature- and field dependence of the bulk pinning force[59] in its most recent form[30] to yield:

$$J_c(\mu_0 H, T) = \frac{Const}{\mu_0 H} \times [\mu_0 H_{c2}(T)]^2 \times h^{1/2}(1-h)^2, \qquad (4)$$

where $h = \mu_0 H / \mu_0 H_{c2}(T)$. Increasing the critical field for $J_c$ scaling from 24.9 to 26.9 T at $T = 4.2$ K thus would result in a rise in non-Cu $J_c(12\ T, 4.2\ K)$ from 2250 A/mm² to 2883 A/mm². This indicates that the non-Cu area in this PIT wire carries ~78 % of what would hypothetically be achievable. The difference with the value found by Cooley et al. (60 %)[30] is due to different area normalization for $J_c$.

It should be noted that a Kramer extrapolated $\mu_0 H_K(4.2\ K) = 24.9$ T arises from a MDG fit using $\mu_0 H_{c2}(0) = 27.7$ T and $T_c(0) = 17.6$ K. These values are a little smaller than for the lower resistive transition at 1 % normal state resistance which yielded $\mu_0 H_{c2}(0) = 28.3$ T and $T_c(0) = 17.8$ K as fitted values. This indicates that the values required for critical current density scaling lie slightly below the range of transitions that are detected in the resistive characterizations. This is indicated in the inset in Figure 8, where the highest- and lowest detected $\mu_0 H_{c2}(T)$ transitions are plotted together with the transition that is appropriate for $J_c$ scaling.

### C. The use of the MDG description

The overall accuracy of the MDG description (2) is demonstrated in the normalized plot in Figure 9. All of our resistively measured data are well described by the MDG fit, independent of whether 1, 10, 50, 90, or 99 % is applied as the criterion for defining $\mu_0 H_{c2}$. Included are resistive measurements on the samples from Table I as well as additional sample material[58], magnetic data using the onset of superconducting behavior[27-29, 58], literature data that has been measured resistively and with RF techniques[1-4] and Kramer extrapolated critical field data resulting from transport $J_c$ characterizations[40, 56]. The route that was followed in order to arrive at Figure 9 is that all $\mu_0 H_{c2}(T)$ data points for a specific sample were least squares fitted with the MDG equation and the resulting values for $\mu_0 H_{c2}(0)$ (calculated from $T_c(0)$ and $D$) and $T_c(0)$ were used as normalization parameters. The only deviations occur very close to $T_c(0)$ in a few of the samples, but the overall shape of the normalized phase boundary is strikingly similar.

The MDG description clearly is efficient in describing the $H$-$T$ phase transition for $Nb_3Sn$ independent of the applied criterion, strain state or sample layout, as is visible from Figure 9. It has the advantage of using only two fitting parameters although the physical meaning of these two parameters can be argued. This description was originally intended to describe dirty, weak coupling superconductors, uses a spherical Fermi surface approximation and does not take into account paramagnetic limiting or spin orbit scattering. The inhomogeneities that are present in practical wires, combined with the lack of accurate A15 resistivity data however rule out rigorous connections to microscopic theories. For theoretical connections to the fundamental parameters, as well as for corrections of the applied approximations, physically more exact descriptions have been developed[48-52] and applied[1, 17, 53, 54]. These more exact descriptions all result, however, in an increased number of fitting parameters, the details of which are impossible to obtain for real, inhomogeneous wire samples. We therefore propose the use of the more simple MDG fit as an acceptable alternative for empirical descriptions of wire data. The accuracy of the description is sufficiently high for reliable estimates of high field behavior from measured low field data up to about 15 T, a field that can be easily obtained with standard superconducting laboratory magnets. This conclusion is important for application to practical scaling relations, where many discussions of what is the optimal function for scaling the field temperature phase boundary have taken place. Figure 9 makes a convincing statement that a simple generalized function is valid for all investigated Nb$_3$Sn.

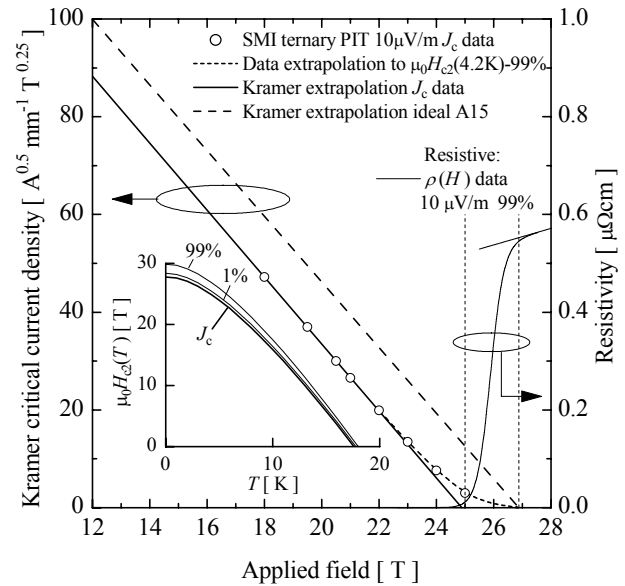

Figure 8: Kramer plot derived from high field transport $J_c$ measurements at 4.2 K with a $10^{-5}$ V/m criterion on a small diameter Ti-6Al-4V barrel for a ternary PIT wire, reacted for 64 hours at 675 °C. In comparison, the small current resistive transition at 4.2 K on a Ti-6Al-4V mounted resistive sample (reacted together with the barrel sample) is also included in the plot and the $10^{-5}$ V/m and 99 % normal state resistance points of this transition are indicated. The $10^{-5}$ V/m point of the resistive transition approximately coincides with the measured $J_c$ point and the onset of the transition, is close to the value for $\mu_0 H_K(4.2\ K)$ from the transport $J_c$ data. The $J_c$ data extrapolation (short dots) assumes a gradual reduction to $\mu_0 H_{c2}$-99 %(4.2 K). The ideal Kramer line (long dots) assumes a hypothetical ideal A15 layer with perfectly homogeneous properties equal to the measured $\mu_0 H_{c2}(T)$-99 % phase transition. A similar hypothetical Kramer line can be simulated from measured Sn gradient profiles as was recently published[30]. The inset depicts the field-temperature boundary that is required for critical current density scaling.

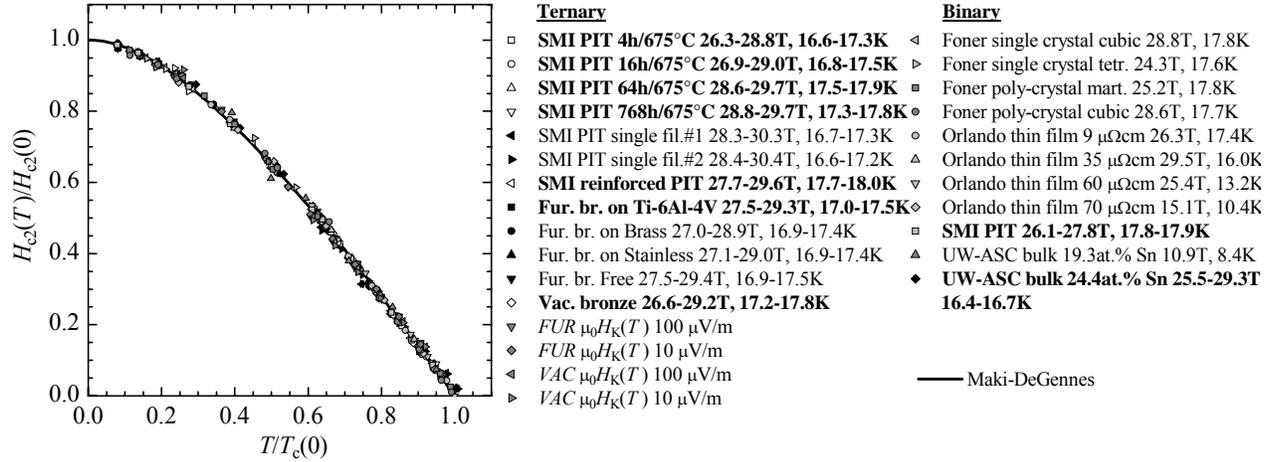

Figure 9: Normalized plot for available $\mu_0 H_{c2}(T)$ data, including data taken from the literature, demonstrating that the shape of the normalized *H-T* phase boundary is identical for all included Nb-Sn phases. The samples in bold are presented in this publication. The *H-T* data were first fitted with the Maki-DeGennes equation and the resulting $\mu_0 H_{c2}(0)$ and $T_c(0)$ were used as normalization parameters. The normalization is valid for 1, 10, 50, 90 and 99 % normal state criteria. The data from the literature uses 50 % (Orlando) and 50-90 % (Foner) criteria. The normalization also holds for Kramer extrapolated critical fields using a 10 or 100 μV/m criterion. The numbers behind the sample names indicate the values for $\mu_0 H_{c2}(0)$ and $T_c(0)$.

The stronger reduction in transition width in $\mu_0 H_{c2}(0)$ compared to $T_c(0)$ implies a change in slope, or in terms of the MDG equation a change in *D* with Sn content. Expansion of the MDG equation and taking the derivative when $T \to T_c$ yields[46]:

$$\left(\frac{d\mu_0 H_{c2}(T)}{dT}\right)_{T=T_c} = -\frac{4\phi_0 k_B}{\pi^2 \hbar D}. \quad (5)$$

Expanding for $T \to 0$ and combining with (2) yields:

$$\mu_0 H_{c2}(0) = \frac{1}{D}\frac{1.76\phi_0 k_B}{h}T_c = -\frac{1.76\pi}{8}T_c\left(\frac{d\mu_0 H_{c2}(T)}{dT}\right)_{T=T_c} \quad (6)$$

We can summarize Devantay's $T_c$(at. % Sn) data on homogeneous bulk samples[16, 19] using a linear fit as usually applied[16, 29, 30] or more accurately with:

$$T_c(at.\%Sn) = \frac{-12.3}{1+\exp\left(\frac{at.\%Sn - 22}{0.9}\right)} + 18.3. \quad (7)$$

We can additionally introduce a function that summarizes the available $\mu_0 H_{c2}$(at. % Sn) data[19, 29]:

$$\mu_0 H_{c2}(at.\%Sn) = -10^{-30}\exp\left(\frac{at.\%Sn}{0.348}\right) + 5.77 at.\%Sn - 107, \quad (8)$$

which includes a $\mu_0 H_{c2}(T)$ suppression near the stoichiometric Sn concentrations due to a cubic to tetragonal phase transition, and combine these with (2) and (6) to fix the *H-T* phase boundary as a function of Sn content if we neglect strain influences.

**VI CONCLUSIONS**

An experimental study has been made of the field-temperature phase boundary in the present generation of optimized but still inhomogeneous wires. The highest detected upper critical fields at zero temperature and the critical temperature at zero field are remarkably constant for all investigated ternary wires and amount to $29.5 \pm 0.3$ T and $17.8 \pm 0.3$ K respectively.

Reduction of wire inhomogeneities by extended heat treatment times was shown by mapping the transition breadth reduction and was demonstrated to be qualitatively the same for resistive and magnetic characterization techniques. The reduction of breadth is attributed to Sn enrichment of the A15 layers in the wires.

The Maki-DeGennes function has proven to yield an excellent description for the field-temperature behavior of Nb$_3$Sn for the entire range of investigated conductors as well as for the thin film and poly- and single crystal data from the literature. In normalized form all $\mu_0 H_{c2}(T)$ dependencies follow one single description, independent of composition, strain state, sample layout or transition criterion. The function also fits the Kramer extrapolated critical fields.

**ACKNOWLEDGMENT**

The authors would like to thank W.A.J. Starch for manufacturing the probe, L. Balicas and B. Brandt for their help with the high field measurements, A. Gurevich and B. ten Haken for valuable discussions and the NHMFL for the use of their 30 T high field system and equipment. Also JAERI Japan and the ITER Europe home team are greatly acknowledged for making the Furukawa and Vacuumschmelze conductors used in this research available. We also would like to thank Shape Metal Innovation, The Netherlands, for donating the PIT wires used in this research. This work was supported by the US Department of Energy, Division of High Energy Physics (DE-FG02-91ER40643), and also benefited from NSF-MRSEC (DMR-9632427) supported facilities.